\documentclass[sigconf, natbib=false]{acmart}
\AtBeginDocument{%
  }

\setcopyright{acmlicensed}
\copyrightyear{2026}
\acmYear{2026}
\acmDOI{XXXXXXX.XXXXXXX}
\acmConference[HPCAsia'26]{The International Conference on High Performance Computing in Asia-Pacific Region}{January 26-29, 2026}{Osaka, Japan}
\acmISBN{978-1-4503-XXXX-X/2018/06}

\RequirePackage[
  datamodel=acmdatamodel,
  style=acmnumeric,
  ]{biblatex}

\usepackage{graphicx}
\usepackage[flushleft]{threeparttable}
\usepackage{tabularx}

\usepackage{amsmath}
\usepackage{tikz}
\usetikzlibrary{patterns}
\usepackage{pgfplots}
\pgfplotsset{compat=1.18}
\usepackage[capitalise]{cleveref}

\usepackage{tabularray}

\usepackage{makecell}

\usepackage{wrapfig}
\definecolor{violet}{rgb}{0.58, 0.0, 0.83}
\usepackage{multirow}

\usepackage{pifont} 
\usepackage{adjustbox}

\definecolor{matplotlibBlue}{HTML}{1f77b4} 
\definecolor{matplotlibOrange}{HTML}{ff7f0e} 
\definecolor{matplotlibGreen}{HTML}{2ca02c} 
\definecolor{matplotlibRed}{HTML}{d62728} 
\definecolor{matplotlibPurple}{HTML}{9467bd} 
\definecolor{matplotlibBrown}{HTML}{8c564b} 
\definecolor{matplotlibPink}{HTML}{e377c2} 
\definecolor{matplotlibGrey}{HTML}{7f7f7f} 
\definecolor{matplotlibYellow}{HTML}{bcbd22} 
\definecolor{myblue}{RGB}{0,114,178}

\usepackage{siunitx} 
\usepackage{xspace} 
\newcommand{\galexi}{GA\-L{\AE}\-XI\xspace}

\begin{document}

\title{Harvesting energy consumption on European HPC systems: Sharing Experience from the CEEC project}


\author{Kajol Kulkarni}
\affiliation{%
  \institution{Friedrich-Alexander-University}
  \city{Erlangen-Nürnberg}
  \country{Germany}}
\email{kajol.kulkarni@fau.de}
\orcid{0009-0009-6726-8005}

\author{Samuel Kemmler}
\affiliation{%
  \institution{Bundesanstalt für Materialforschung und -prüfung}
  \city{Berlin}
  \country{Germany}}
\email{samuel.kemmler@fau.de}
\orcid{0000-0002-9631-7349}

\author{Anna Schwarz}
\affiliation{%
  \institution{University of Stuttgart}
  \city{Stuttgart}
  \country{Germany}}
\email{schwarz@iag.uni-stuttgart.de}
\orcid{0000-0002-3181-8230}

\author{G\"ul\c{c}in Gedik}
\orcid{0009-0001-8380-1289}
\author{Yanxiang Chen}
\affiliation{%
  \institution{Ume\aa{} University}
  \city{Ume\aa{}}
  \country{Sweden}}
\email{ychen@cs.umu.se}
\orcid{0009-0003-5512-254X}

\author{Dimitrios Papageorgiou}
\orcid{0009-0009-0854-8971}
\author{Ioannis Kavroulakis}
\affiliation{%
  \institution{Aristotle University of Thessaloniki}
  \city{Thessaloniki}
  \country{Greece}}
\email{{dpapageor,ikkavroul}@meng.auth.gr}
\orcid{0009-0004-9841-9615}

\author{Roman Iakymchuk}
\affiliation{%
  \institution{Ume\aa{} University and\\ Uppsala University}
  \city{Ume\aa{}}
  \country{Sweden}}
\email{riakymch@cs.umu.se}
\orcid{0000-0003-2414-700X}

\renewcommand{\shortauthors}{Kajol Kulkarni et al.}

\begin{abstract}
Energy efficiency has emerged as a central challenge for modern high-performance computing (HPC) systems, where escalating computational demands and architectural complexity have led to significant energy footprints. This paper presents the collective experience of the EuroHPC JU Center of Excellence in Exascale CFD (CEEC) in measuring, analyzing, and optimizing energy consumption across major European HPC systems. We briefly review key methodologies and tools for energy measurement as well as define metrics for reporting results. Through case studies using representative CFD applications (waLBerla, FLEXI/GALÆXI, Neko, and NekRS), we evaluate energy-to-solution and time-to-solution metrics on diverse architectures, including CPU- and GPU-based partitions of LUMI, MareNostrum5, MeluXina, and JUWELS Booster. Our results highlight the advantages of accelerators and mixed-precision techniques for reducing energy consumption while maintaining computational accuracy. Finally, we advocate the need to facilitate energy measurements on HPC systems in order to raise awareness, teach the community, and take actions toward more sustainable exascale computing.
\end{abstract}





\maketitle

\section{Introduction}



The continuous growth in computational demand and system complexity in large-scale high-performance computing (HPC) systems \cite{JD14,Reed2015ExascaleCA,etp4hpcsra5} has made energy consumption a critical design and operational concern. Modern supercomputers consist of thousands, sometimes millions, of interconnected processing units working concurrently, which results in significant power requirements and environmental impact. This growing energy footprint not only drives operational costs and limits scalability but also challenges the sustainability objectives of next-generation exascale systems. Consequently, the HPC community has begun to experience a paradigm shift and to reassess hardware architectures, numerical algorithms, especially those used in fundamental linear algebra operations, and software design, with a view to improving energy efficiency without compromising computational performance.

This shift can be attributed to an increased focus on sustainable computational practices, which align with the Swedish principle of lagom, advocating moderation or `just enough' precision in computation and data storage. In a similar manner, the ETP4HPC Strategic Research Agenda \cite{etp4hpcsra5} further advances this notion through the concept of energy-to-solution, linking it to the success of mixed-precision methods in numerical solvers. 

Despite increasing attention to energy efficiency, the systematic integration of energy-to-solution into conventional HPC performance metrics, such as runtime, scalability, and resource utilization, has not yet been realized. Energy profiling is inherently more complex than traditional time measurement because it depends on hardware counters, privileged system access, and vendor-specific monitoring tools. The resulting diversity of interfaces, measurement granularity, and data interpretation complicates reproducibility and cross-system benchmarking. Additionally, legacy scientific applications, built and honed over decades, pose more challenges. Extensive, optimized codebases may lack the necessary flexibility and instrumentation for energy-awareness integration. Although energy efficiency is well recognized, its actual implementation in production-grade systems is still limited.

To bridge this gap, this paper presents an initiative from the EuroHPC Joint Undertaking (JU) Center of Excellence in Exascale CFD (CEEC) aimed at empowering HPC users and developers to understand, measure, and reduce energy consumption effectively. This work is a continuation of the CEEC Best Practice Guide~\cite{ceec-bpg}. We found that energy measuring procedures vary widely among systems and are typically more complicated than time-to-solution measurement utilizing profiling APIs. Lack of consistent methodology and restricted user access to energy counters impede application-level energy evaluation. Even in limited situations, non-privileged APIs, software energy libraries, and workload-level estimators can provide significant energy monitoring.

In this paper, we present an overview of energy measurements on European HPC systems, including CEEC's experience, tools, and lessons learned. Remainder of paper is arranged as follows: \cref{sec:how-to-measure} covers energy consumption measurement and categorization options. This provides information about hardware measuring techniques in different architectures and tools and frameworks for these counters. With CEEC examples, \cref{sec:walberla,sec:flexi,sec:neko,sec:nekrs} describes our energy measuring methodologies on European systems. Finally, \cref{sec:conclusions} concludes the article. 

\section{How to measure energy consumption}
\begin{figure}[!ht]
\hspace*{-2mm}\includegraphics[width=0.52\textwidth]{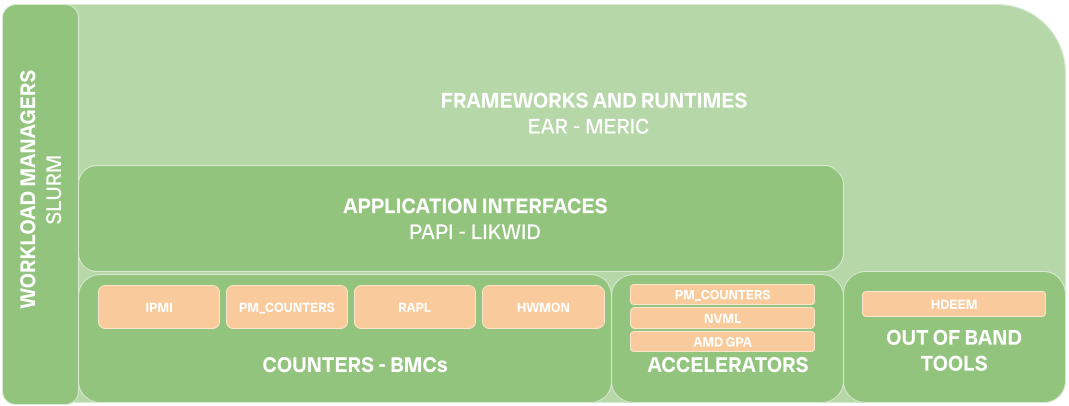}
\caption{Hierarchical relation between energy measurement methods.
}
\label{fig: hierarhcy-software}
\end{figure}

In this section, we briefly describe best practices and existing tools, as well as share our own experience with reliably measuring energy on different platforms. \cref{fig: hierarhcy-software} provides a bird eye view of the tools and frameworks that are going to be examined, and it presents the spectrum from physical methods to software techniques.

\label{sec:how-to-measure}
\subsection{Hardware Counters}
\subsubsection{RAPL} Running Average Power Limit (RAPL) \cite{rapl} is a set of tools introduced by Intel to enable fine-grained monitoring and control of processor power consumption. RAPL utilizes Model Specific Registers (MSRs) that are specific to each model processors to provide an interface for managing and observing multiple power domains, such as the processor core, package, DRAM, and system-level components. These registers can be programmed to enforce power consumption constraints, ensuring compliance with thermal design limits or maintaining a desired balance between power efficiency and performance. MSR interfaces in each RAPL domain provide Watt-based power limitation, report energy usage in model-specific units (usually microjoules), and highlight performance implications from power capping. RAPL enables hierarchical control across its domains at the socket level. It can thus function both as a power management mechanism and as a power measurement tool \cite{7284406}.
\subsubsection{IPMI} The Intelligent Platform Management Interface (IPMI) \cite{ipmispec} is a standardize collection of interfaces that enables the management and monitoring of computer systems independently of the host operating systems, utilizing Baseboard Management Controllers (BMCs). These interfaces provides access to node power consumption monitoring. It was demonstrated in \cite{quantativecomp} that, despite the low sampling rate of IPMI, the power measurement data remains sufficiently accurate. However, it should be noted that energy consumption measurements derived from these data may exhibit reduced precision. Moreover, the restricted sampling frequency is a significant limitation, which is primarily ascribed to the substantial operational overhead intrinsic to IPMI communication.
\subsubsection{NVIDIA GPUs}HPC systems are inherently heterogeneous, posing challenges for accurate hybrid power measurements. As noted in the SCALABLE Deliverable \cite{d2.3}, CPUs and discrete GPUs lack a common power interface, except in integrated architectures (e.g., RAPL). Consequently, measurements must be combined from multiple sources. For NVIDIA GPUs, the NVML API [39] provides energy data since the last driver reload, including memory and related circuitry, with accuracy and sampling rates varying by architecture. Whereas, in NVIDIA Grace SoCs, both CPU and GPU power metrics are directly available.
\subsubsection{PM\_Counters} On Cray systems such as LUMI and Dardel at KTH, power monitoring is supported for both CPU-only and hybrid CPU–GPU configurations \cite{crayx30powermonitorandmanage}. Node-level counters that provide real-time power consumption and total energy usage (in Joules) are available to non-privileged users through {\tt /sys/cray/pm\_counters}, enabled by {\tt cray-pat}. These counters update atomically every 100 ms \cite{crayx30}, necessitating synchronization between energy readings and their corresponding measurements for consistency. A ‘freshness’ counter is provided to ensure data integrity by being read before and after sampling. The same measurements are also available through Score-P and the Vampir performance analysis tools.

\Cref{tab:summary_counters} summarizes the variety of the counters presented with unique advantages such as granularity, metric, and overhead.

\begin{table*}
\caption{Summarizing Properties of Hardware Counters.}
\label{tab:summary_counters}
\centering
\begin{adjustbox}{width=0.76\textwidth}
\begin{threeparttable}
\begin {tabular}{|c|c|c|c|c|c|c|}
\hline
\textbf{Type} & \textbf{Domains} & \textbf{Granularity} & \textbf{Measurement Type}  & \textbf{Metric} & \textbf{Vendor Specific} & \textbf{Overhead}\tnote{1}\\ 
\hline 
RAPL &  \makecell{Socket(PP0) \\ All cores \\ DRAM \\ System \\ Integ. GPU \\ (AMD) Per Core  }  & 1 ms  & Counter & mili-Joules & \ding{51} & Low\\ 
\hline 
NVML & Socket & -\tnote{2}  & Counter & mili-Joules  & \ding{51} & - \\
\hline 
IPMI & Node & -\tnote{3} & BMC & Watt & \ding{55} & High\\
\hline 
PM\_COUNTERS & \makecell{Node \\ Network Card \\ GPU } & 100 ms & Counter  & Joules  & \ding{51} & Medium\\
\hline 
HDEEM & \makecell{Blade \\ Node \\ DRAM } & 1 ms &\makecell{FPGA \\ BMC } & Watt  & \ding{55}  & Low \\
\hline 
\end{tabular}
  \begin{tablenotes}
    \item[1] Short measurement intervals.
    \item[2] Architecture dependant.
    \item[3] BMC sensor dependant.
  \end{tablenotes}
\end{threeparttable}
\end{adjustbox}
\end{table*}

\subsection{Tools and Frameworks to measure Energy Consumption}
Building on the hardware counters described above, software-level measurement frameworks such as Perf ~\cite{perfwiki}, PAPI~\cite{powermonitorwithpapi}, LIKWID~\cite{likwid}, and HDEEM~\cite{hdeempaper} provide accessible means for collecting, analyzing, and correlating energy data with application performance. Each tool, when combined with these counters, supports distinct research objectives, such as varying spatial and temporal granularity. Beyond individual measurement tools, there are ready to use runtime systems and APIs, including Energy Aware Runtime (EAR) \cite{earbsc}, MERIC ~\cite{mericpaper}, LLview \cite{LLview2023}, and SLURM are designed to measure and optimize energy consumption. This section presents an overview of frameworks used in this paper.

\subsubsection{SLURM Workload Manager}
SLURM \cite{slurm} provides a unified and user-friendly interface for application-wide energy measurements, supporting various socket-level infrastructures such as {\tt IPMI}, {\tt RAPL} or {\tt PM\_Counters}. It is a common practice to deploy this system with root privileges.  In order to enable energy accounting and monitoring, it is necessary to activate the appropriate plug-in in the slurm.conf configuration file. This configuration file also specifies the underlying measurement method. For instance, in the EuroHPC JU system LUMI, this plug-in employs {\tt PM\_Counters} from HPE Cray Systems, facilitated by the BMC, thereby providing node-level data. Likewise, on Meluxina, SLURM is also used to measure node-level energy consumption, relying on IPMI counters for data collection. Consequently, it is imperative to ensure process isolation and minimize OS noise to facilitate accurate measurements. Additionally, SLURM offers an External Sensors Plug-in for the purpose of collecting out-of-band energy and temperature data from external monitoring systems such as Nagios or wattmeters.
\subsubsection{EAR}
Energy Aware Runtime (EAR)~\cite{earbsc} is a toolset for monitoring and optimizing system power and job energy usage. It enables runtime energy optimization, cluster management, and power capping without modifying applications. To enable energy optimization, EAR uses its runtime library~\cite{earpaper}, which dynamically selects the optimal CPU frequency according to the energy policy and application behavior. Note that EAR requires to be installed with root privileges on the Linux kernel.

\subsubsection{LLview}
\label{sec:subsec:subsubsec-LLview}
LLview~\cite{LLview2023} is a cluster monitoring tool that interfaces with resource managers and schedulers. Its Job Reporting module provides detailed, near real-time insights into running jobs, collecting data from compute nodes and GPUs with minimal overhead. Metrics tracked include GPU usage, memory, power, and temperature, updated every minute via NVIDIA’s DCGM, while additional daemons collect extra data from compute nodes. 
A web portal not only links performance metrics to specific jobs but also allows users to access tables with aggregated performance information, timeline graphs detailing key metrics over the course of a job, and reports in both interactive and PDF formats. Additionally, this efficient interface {\em updates performance metrics every minute} to provide real-time data, ensuring minimal system overhead. For instance, the values are obtained for each GPU from NVIDIA’s Data Centre GPU Manager {\tt dcgm} every minute.

\subsection{Metrics}
It is essential to establish an energy measurement and integrate it into the conventional HPC performance measurements, including time-to-solution and its related measures. Apart from measuring typical metrics such as energy consumption or energy-to-solution, several other metrics can provide deeper insights into system performance. This section presents an overview of the most relevant metrics.

\subsubsection{Energy Ratio}
Energy ratio can be defined as the ratio of energy consumed by a parallel simulation to that of a baseline simulation executed with the minimum number of ranks:
\begin{equation}
  \text{Energy Ratio} =                                                                  \frac{\text{Energy}_{P}}{\text{Energy}_{P_{0}}},
  \label{eq:energy_eff}
\end{equation}
where $\text{Energy}_{P}$ is the energy required for a simulation with $P$ ranks and $\text{Energy}_{P_{0}}$ is the energy required for a simulation with the minimum number of ranks $P_{0}$. A higher energy ratio signifies reduced energy efficiency, as a greater amount of energy is required to complete the same simulation workload. As parallel efficiency decreases, the energy ratio correspondingly increases, approximately in proportion to the loss of parallel efficiency. 

\subsubsection{Energy-normalized Performance Index (EPID)}
\label{sec:epid}
With the growing use of accelerator-based architectures, rank-based performance comparisons between CPU and GPU systems are less meaningful due to differences in computational power per rank. To address this, an energy-normalized performance index (PID) has been defined~\cite{kempf2024galaexi} to incorporate total energy consumption into the performance evaluation, which is defined as
\begin{equation}
  \text{EPID} = \frac{\mathrm{walltime} \times \mathrm{power}}{\mathrm{\#RK}\text{-}\mathrm{stages} \times \mathrm{\#DoF}} =                                                                   \underbrace{\frac{\mathrm{power}}{\mathrm{\#ranks}}}_{P_{\text{rank}}}\;\times\;\mathrm{PID},
\end{equation}
representing the energy required to compute the time update for a single degree of freedom (DoF) for one Runge–Kutta stage (RK for short) on the specific computing hardware; the metric is also valid for other methods, even iterative.
As such, the EPID is the PID normalized by the specific power required per rank, $P_{\text{rank}}$. A lower PID indicates better performance.



\section{waLBerla}
\label{sec:walberla}
Within the scope of the CEEC project, the \textsc{waLBerla} multiphysics framework \cite{bauer2021walberla} is employed to develop a fully resolved, coupled fluid-particle numerical model, referred to as lighthouse case 4 (\textsc{LHC4})\footnote{More details on lighthouse cases are on \url{https://ceec-coe.eu/lighthouse-cases/}.}. This model is designed to investigate the phenomenon of piping erosion, which poses a significant threat to geotechnical structures such as offshore wind turbine foundations and dams. 
The numerical approach combines the lattice Boltzmann method for the simulation of the fluid phase with the discrete element method to model the granular soil. The implementation details of this coupled framework have been presented in \cite{kemmlerEfficiencyScalabilityFullyresolved2025}, while the numerical model itself, including its validation, is discussed comprehensively in \cite{kemmlerFullyresolvedMicromechanicalSimulation2025a}. 
Each time step of the fully resolved coupled fluid-particle simulation comprises several components, including the fluid dynamics computation, the coupling between the fluid and particle phases, boundary handling, and particle dynamics. Among these, three modules dominate the computational cost:
\begin{itemize}
    \item \textit{Fluid:} simulates the fluid dynamics (stencil-based kernel).
    \item \textit{Mapping:} maps particle information onto the fluid grid (square root operation for every cell).
    \item \textit{Reduction:} aggregates hydrodynamic forces acting from the fluid on the particles (global reduction operation). 
\end{itemize}
The algorithms underlying these components are described in detail in \cite{kemmlerTowardsExascaleSimulations2026}. 
The following sections present a detailed analysis of the energy consumption of the \textsc{LHC4} model on a single node of the LUMI supercomputer, considering both CPU and GPU partitions. The analysis begins with an overview of the simulation setup and domain decomposition strategy, followed by a quantitative evaluation of time-to-solution and energy-to-solution for the key computational components.

\subsection{Simulation Setup}
\begin{figure}
    \centering
    \includegraphics[height=0.26\textheight]{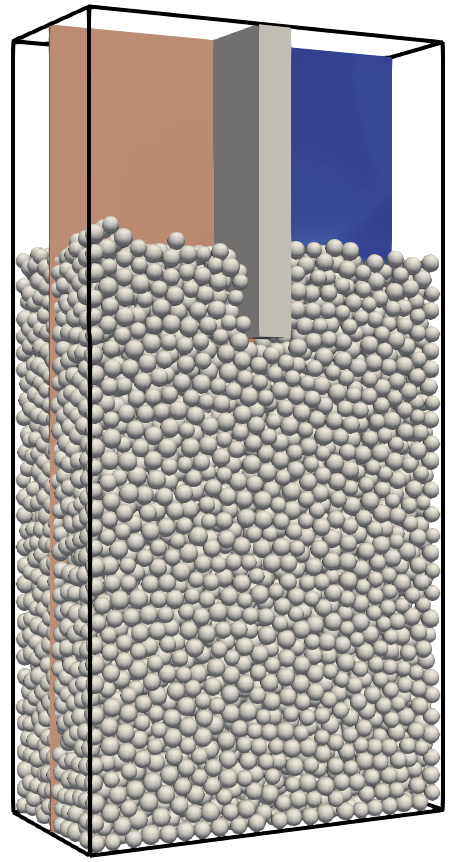}
    \caption{Three-dimensional numerical setup of LHC4, \textsc{waLBerla} \cite{kemmlerTowardsExascaleSimulations2026}.}\label{fig:simulation_setup}
\end{figure}

The simulation domain, illustrated in \cref{fig:simulation_setup}, is a rectangular box with dimensions $448 \times 224 \times 896$, corresponding to a total of $89{,}915{,}392$ fluid cells. The domain contains $8{,}796$ particles and a large box-shaped geometry, representing a simplified section of granular soil through which fluid flows.
To enable large-scale parallel execution, the simulation domain is decomposed into multiple uniform subdomains, i.e., blocks. Each block is assigned to a single MPI process. On CPU nodes, the number of MPI processes is equal to the number of available cores, i.e., 128 on a LUMI-C node. On GPU nodes, the number of MPI processes corresponds to the number of Graphics Compute Dies (GCDs), which is eight per LUMI-G node, organized into four GPUs per node. 
An overview of the domain decomposition for both partitions is provided in~\cref{tab:domain_partitioning}. Although the number of blocks differs between CPU and GPU configurations, both setups represent exactly the same total number of fluid cells, thereby ensuring identical problem sizes across hardware platforms. This enables a direct comparison of time-to-solution and energy-to-solution between the two architectures.
\begin{table}
    \centering
    \caption{Domain partitioning on LUMI.}
    \begin{tabular}{|c|c|c|c|c|c|c|c|}
        \hline
        \textbf{\shortstack{\rule{0pt}{2ex}Partition\\ \phantom{dummy}}} & \textbf{\shortstack{\rule{0pt}{2ex}Cores/GCD\\ per node}} & \multicolumn{3}{c|}{\textbf{\shortstack{\rule{0pt}{2ex}MPI\\ processes}}} & \multicolumn{3}{c|}{\textbf{\shortstack{\rule{0pt}{2ex}Fluid cells\\ per process}}} \\ 
        & & \textbf{x} & \textbf{y} & \textbf{z} & \textbf{x} & \textbf{y} & \textbf{z}\\
        \hline
        LUMI-C & 128 & 4 & 2 & 16 & 112 & 112 & 56 \\ \hline
        LUMI-G & 8 & 2 & 1 & 4 & 224 & 224 & 224 \\ \hline
    \end{tabular}
    \label{tab:domain_partitioning}
\end{table}

\subsection{Energy Consumption}
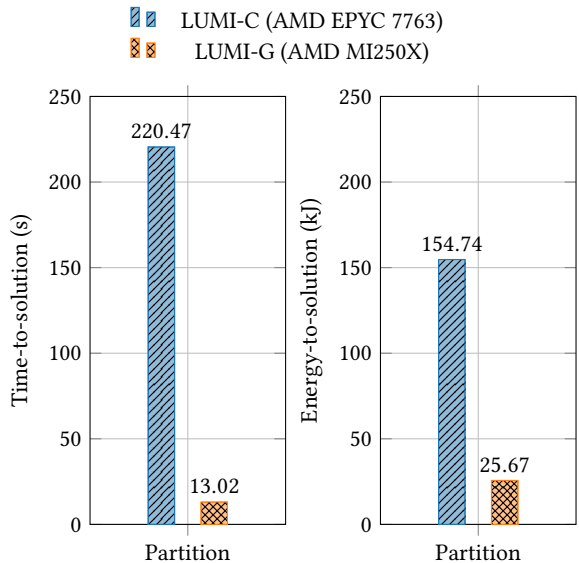
\begin{figure}
    \centering
    \begin{center}
        \ref{combined_legend}
    \end{center}
    \begin{tikzpicture}
        \begin{axis}[
            ybar,
            ylabel={Time-to-solution (s)},
            symbolic x coords={Partition},
            xtick=data,
            ymin=0,
            ymax=250,
            legend style={draw=none, legend columns=1, column sep=2ex},
            legend entries={LUMI-C (AMD EPYC 7763), LUMI-G (AMD MI250X)},
            legend to name=combined_legend,
            nodes near coords,
            grid,
            x=1.3cm,
        ]
        \addplot[draw=matplotlibBlue,fill=matplotlibBlue!50, bar shift=-10pt, postaction={pattern=north east lines}] plot coordinates {
            (Partition, 220.47)
        };
        \addplot[draw=matplotlibOrange,fill=matplotlibOrange!50, bar shift=10pt, postaction={pattern=crosshatch}] plot coordinates {
            (Partition, 13.02)
        };
        \end{axis}
    \end{tikzpicture}
    \begin{tikzpicture}
        \begin{axis}[
            ybar,
            ylabel={Energy-to-solution (kJ)},
            symbolic x coords={Partition},
            xtick=data,
            ymin=0,
            ymax=250,
            nodes near coords,
            grid,
            x=1.3cm,
        ]
        \addplot[draw=matplotlibBlue,fill=matplotlibBlue!50, bar shift=-10pt, postaction={pattern=north east lines}] plot coordinates {
            (Partition, 154.74)
        };
        \addplot[draw=matplotlibOrange,fill=matplotlibOrange!50, bar shift=10pt, postaction={pattern=crosshatch}] plot coordinates {
            (Partition, 25.67)
        };
        \end{axis}
    \end{tikzpicture}
    \caption{Time-to-solution (left) and energy-to-solution (right) on a single node run of LHC4, \textsc{waLBerla}.}
\label{fig:walberla_energy}
\end{figure}

\begin{figure}
    \centering
    \begin{center}
        \ref{combined_legend_2}
    \end{center}
    \resizebox{0.45\textwidth}{!}{
    \begin{tikzpicture}
        \begin{axis}[
            ybar,
            ylabel={Time-to-solution (s)},
            symbolic x coords={Fluid, Mapping, Reduction},
            xtick=data,
            enlarge x limits=0.25,
            width=0.35\textwidth,
            height=0.4\textheight,
            ymin=0,
            ymax=150,
            ytick={0,30,...,150},
            legend style={draw=none, legend columns=1, column sep=2ex},
            legend to name=combined_legend_2,
            nodes near coords,
            grid,
        ]
        \addplot[draw=matplotlibBlue,fill=matplotlibBlue!50,bar shift=-8pt,postaction={pattern=north east lines}] plot coordinates {
            (Fluid, 140.65)
            (Mapping, 22.39)
            (Reduction, 3.96)
        };
        \addplot[draw=matplotlibOrange,fill=matplotlibOrange!50,bar shift=8pt,postaction={pattern=crosshatch}] plot coordinates {
            (Fluid, 6.70)
            (Mapping, 1.34)
            (Reduction, 1.39)
        };
        \legend{LUMI-C (AMD EPYC 7763), LUMI-G (AMD MI250X)}
        \end{axis}
    \end{tikzpicture}
    \begin{tikzpicture}
        \begin{axis}[
            ybar,
            ylabel={Energy-to-solution (kJ)},
            symbolic x coords={Fluid, Mapping, Reduction},
            xtick=data,
            enlarge x limits=0.25,
            width=0.35\textwidth,
            height=0.4\textheight,
            ymin=0,
            ymax=150,
            ytick={0,30,...,150},
            nodes near coords,
            grid,
        ]
        \addplot[draw=matplotlibBlue,fill=matplotlibBlue!50,bar shift=-10pt,postaction={pattern=north east lines}] plot coordinates {
            (Fluid, 110.60)
            (Mapping, 19.87)
            (Reduction, 4.42)
        };
        \addplot[draw=matplotlibOrange,fill=matplotlibOrange!50,bar shift=10pt,postaction={pattern=crosshatch}] plot coordinates {
            (Fluid, 16.00)
            (Mapping, 5.85)
            (Reduction, 5.48)
        };
        \end{axis}
    \end{tikzpicture}
    }
    \caption{Time-to-solution (left) and energy-to-solution (right) on a single node run for the dominating simulation modules of LHC4, \textsc{waLBerla}.}
    \label{fig:walberla_energy_modules}
\end{figure}
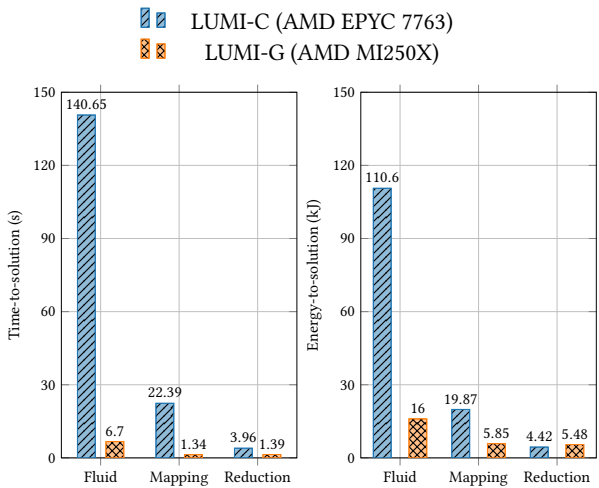

Energy consumption measurements were obtained after the completion of each job using the SLURM workload manager via the command \texttt{sacct --format="JobID,ConsumedEnergy" -j <jobId>}. To assess the energy usage of individual modules of the multiphysics simulation, separate runs were conducted in which only one module was active while the others were disabled. 
The energy metrics were recorded at the node level to ensure accuracy, as reliable readings can only be obtained when a compute node is allocated exclusively to a single job. Each simulation consisted of 500 time steps.
\cref{fig:walberla_energy} presents the measured time-to-solution and energy-to-solution for the entire multiphysics simulation, while \cref{fig:walberla_energy_modules} shows corresponding results for the three dominant modules described above. The measurements reveal substantial performance and energy differences across architectures.
As expected, the time-to-solution is significantly lower on the GPU nodes compared to the CPU nodes, both for the complete multiphysics simulation and for each of the three dominant simulation modules. The energy-to-solution follows a similar trend, with GPUs generally demonstrating superior energy efficiency. An exception is observed for the reduction module, where the CPU node slightly outperforms the GPU node in terms of energy consumption. These findings indicate that GPUs offer substantial advantages in both performance and energy efficiency; however, for certain computational kernels, such as reductions, CPUs may still exhibit favorable characteristics. These findings must be assessed in the light of GPU nodes being significantly more expensive than CPU nodes.

\section{FLEXI}
\label{sec:flexi}
In the context of CEEC, the open-source, high-order accurate flow solver FLEXI and its GPU counterpart \galexi~\cite{flexi,kempf2024galaexi,Keim2025} are utilized to examine the shock buffet on a 3D wing under transonic flight conditions, referred to as lighthouse case 1 (LHC1).
Shock buffet leads, among other phenomena, to increased structural fatigue of the wing in the long run, significantly affecting the safety and efficiency of the aircraft.
As a numerical discretization scheme, the discontinuous Galerkin (DG) spectral element method of polynomial order $N \in \mathbb{N}_{>0}$ is utilized combined with a localized low-order finite volume subcell approach to mitigate oscillations near discontinuities, as commonly induced by high-order schemes. 
The fluid phase is governed by the compressible Navier--Stokes--Fourier equations. 
The temporal integration is performed by using explicit, high-order, low-storage Runge--Kutta schemes.
Further details on the numerical method and its validation are presented in~\cite{flexi,kempf2024galaexi,Schwarz2025b}.
The subsequent section provides a comprehensive view on the energy consumption of LHC1, employing a simplified model to reduce computational complexity.
First, the simulation setup and domain decomposition strategy are briefly outlined.
This is followed by a detailed analysis of the energy consumption considering the energy efficiency and EPID, see~\cref{sec:epid}, on multiple nodes as well as the time-to-solution and energy-to-solution on a single node. The first investigations are carried out on the CPU (AMD EPYC 7452) and GPU (NVIDIA A100) partitions of the MeLuxina supercomputer, while the latter are performed on the GPU (NVIDIA H100) and CPU (Intel Sapphire Rapids 8480+) partitions of MareNostrum5 and the GPU (AMD MI250X) partition of LUMI.

\subsection{Simulation Setup}
The simulation domain is a Cartesian box $\Omega\in[0,1]^3$ composed of hexahedral elements, and periodic boundary conditions are prescribed. 
All simulations are initialized with a constant flow state using polynomial degrees of $N=5$ (CPU) and $N=9$ (GPU) and, to mimic LHC1, every second DG element is switched to finite volume (FV). 
FLEXI is parallelized using pure MPI communication, and mesh elements are pre-sorted along a space-filling curve by the open-source mesh generator PyHOPE \cite{pyhope}. The solution data is structured in such a manner that it forms linear memory segments for communication, and is stored using the distributed memory paradigm. Since the computational stencil of the DG scheme is element-local, only the numerical flux between individual elements has to be exchanged. Concurrently, intra-element computation is employed to facilitate efficient latency hiding.


\subsection{Energy Consumption}
\begin{figure}
    \centering
    \begin{center}
        \ref{flexi_leg1_comb}
    \end{center}
    \begin{tikzpicture}
        \begin{axis}[
            width=0.25\textwidth,
            height=0.3\textwidth,
            ymin=0.1,ymax=1.1,
            xmin=0.5*128,xmax=1.5*16384,
            xlabel={\# CPUs},
            xmode=log,
            legend to name=flexi_leg1_comb,
            legend style={draw=none, legend columns=2, column sep=2ex},
            grid,
        ]
        \addplot[draw=matplotlibBlue,thick,mark=o] table {
            128 1.0
            256 0.958177099228558
            512 0.9665892991756624
            1024 0.88092999645358035
            2048 0.8766824390370493
            4096 0.7501082081267116
            8192 0.7820423467709028
            16384 0.7493162433605811
        };
        \addplot[draw=matplotlibOrange,thick,mark=o] table {
            128 1.0
            256 0.7283388448835004
            512 0.6619130745995261
            1024 0.584845398248588
            2048 0.5723132078040553
            4096 0.4033898880967962
            8192 0.29675043188322525
            16384 0.16554736724472344
        };
        \addlegendentry{Parallel efficiency (PID)}
        \addlegendentry{Energy efficiency (EPID)}
        \end{axis}
    \end{tikzpicture}
    \begin{tikzpicture}
        \begin{axis}[
            width=0.25\textwidth,
            height=0.3\textwidth,
            ymin=0.1,ymax=1.1,
            xmin=0.5*4,xmax=1.5*128,
            xlabel={\# Devices},
            xmode=log,
            grid,
        ]
        \addplot[draw=matplotlibBlue,thick,mark=o] table {
            4 1.0
            8 0.9721280526023963
            16 0.9710431756184743
            32 0.9398265421911212
            64 0.934459410529365
            128 0.9329703333950303
        };
        \addplot[draw=matplotlibOrange,thick,mark=o] table {
            4 1.0
            8 0.9998840184366476
            16 0.9988840184366476
            32 0.2450544591898903
            64 0.6119515686340506
            128 0.4968425763781357
        };
        \end{axis}
    \end{tikzpicture}
    \caption{Energy efficiency (EPID) and parallel efficiency (PID) of FLEXI/\galexi on the CPU (left) and GPU (right) partition of MeluXina.}
    \label{fig:flexi_energy_efficiency}
\end{figure}
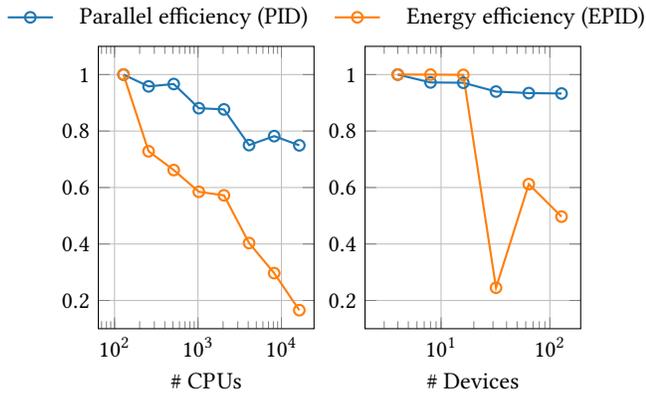

Similarly to \textsc{waLBerla}, energy consumption measurements were obtained after the completion of each job using the SLURM workload manager via the command \texttt{sacct --format="JobID,ConsumedEnergy" -j <jobId>}. 
In order to assess the energy usage of FLEXI for multiple nodes, the energy consumption is measured for a constant number of DoFs per node and a constant number of total DoFs, in a manner analogous to weak and strong scaling, respectively, on the GPU and CPU partitions of the MeLuxina supercomputer.
The number of DoFs is varied by changing the number of elements in the computational domain for a fixed $N$.
The energy and parallel efficiency of FLEXI/\galexi, measured by the EPID and PID, respectively, using \num{6912} DoFs per node (CPU) and \num{2.048e6} DoFs per node (GPU) normalized by the performance on one node (128 CPUs / 4 GPUs), as depicted in~\cref{fig:flexi_energy_efficiency}.
The findings of the study demonstrate that the energy efficiency decreases proportional to the parallel efficiency for an increasing number of CPU ranks but drops significantly faster than the parallel efficiency.
Conversely, the results for the GPU partition demonstrate a different behavior: Up to 16 devices, the energy and parallel efficiency are comparable, while for a larger number of devices, the energy efficiency significantly drops, particularly for 32 devices, compared to its counterpart which remains up to 90\%. 
A direct comparison to the CPU results reveals that as expected the energy efficiency on GPUs is superior to that on CPUs.
The energy efficiency, as defined in (\ref{eq:energy_eff}), times the number of CPUs on one node, and the speedup of FLEXI are depicted in~\cref{fig:flexi_energy_efficiency2} using a constant number of \num{2.81e7} DoFs (CPU) and \num{1.31e8} DoFs (GPU) normalized by the performance on one node.
The results indicate a linear increase in energy consumption with an increasing number of CPUs, with a slight dip at 8 ranks, combined with a nearly superlinear speedup. Conversely, the energy consumption on the GPU partition shows no real trend, only a slight increase can be observed for a larger number of devices. This behavior can be attributed to the fact that GPUs require a certain amount of workload, usually $\approx 2e6$ DoFs~\cite{kempf2024galaexi}, to operate optimally, but here the workload on each device depends on the number of ranks.
A direct comparison of the performance of FLEXI on GPU and CPU partitions (using $1.6e7$ DoFs) reveals that the energy-to-solution and time-to-solution on a single node highlight the significant reduction in runtime and energy consumption when using AMD or NVIDIA GPUs compared to CPUs, cf.~\cref{fig:flexi_energy}, with the former showing the best results for the given configuration.
Overall, the findings indicate that GPU partitions have yielded superior energy-to-solution metrics in comparison to CPU-based systems, particularly for a larger number of ranks, under the condition that the available compute resources are utilized to their maximum potential.

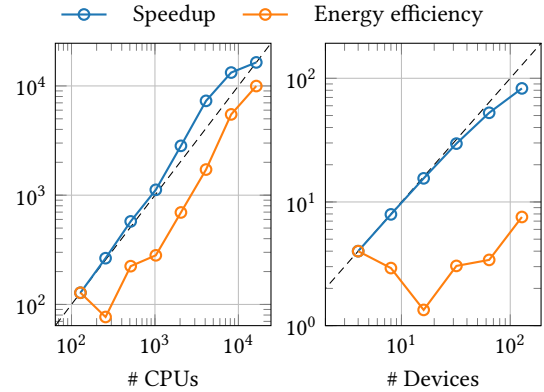
\begin{figure}
    \centering
    \begin{center}
        \ref{flexi_leg3_comb}
    \end{center}
    \begin{tikzpicture}
        \begin{axis}[
            width=0.25\textwidth,
            height=0.3\textwidth,
            domain=0.5:2*16384,
            ymin=0.5*128,ymax=1.5*16384,
            xmin=0.5*128,xmax=1.5*16384,
            xlabel={\# CPUs},
            xmode=log,ymode=log,
            legend to name=flexi_leg3_comb,
            legend style={draw=none, legend columns=2, column sep=2ex},
            grid,
        ]
        \addplot+[mark=none,black,densely dashed, forget plot] (x,x);
        \addplot[draw=matplotlibBlue,thick,mark=o] table {
            128 128.0
            256 265.0481144807167
            512 575.5905707676582
            1024 1119.8407806903883
            2048 2829.630395975976
            4096 7299.708186410572
            8192 13254.43565088047
            16384 16412.398849560912
        };
        \addplot[draw=matplotlibOrange,thick,mark=o] table {
            128 128.0
            256 76.8
            512 224
            1024 281.6
            2048 695.04
            4096 1716.48
            8192 5488.64
            16384 9973.76
        };
        \addlegendentry{Speedup}
        \addlegendentry{Energy efficiency}
        \end{axis}
    \end{tikzpicture}
    \begin{tikzpicture}
        \begin{axis}[
            width=0.25\textwidth,
            height=0.3\textwidth,
            ymin=1.0,ymax=1.5*128,
            domain=0.5:2*128,
            xmin=0.5*4,xmax=1.5*128,
            xlabel={\# Devices},
            xmode=log,ymode=log,
            grid,
        ]
        \addplot+[mark=none,black,densely dashed, forget plot] (x,x);
        \addplot[draw=matplotlibBlue,thick,mark=o] table {
            4 4.0
            8 7.935121841727742
            16 15.518396134343703
            32 29.635015407044172
            64 52.6039089121285
            128 82.83872745406879
        };
        \addplot[draw=matplotlibOrange,thick,mark=o] table {
            4 4.0
            8 2.92
            16 1.34
            32 3.04
            64 3.4
            128 7.56
        };
        \end{axis}
    \end{tikzpicture}
    \caption{Energy efficiency and speedup of FLEXI/\galexi on the CPU (left) and GPU (right) partition of MeluXina.}
    \label{fig:flexi_energy_efficiency2}
\end{figure}
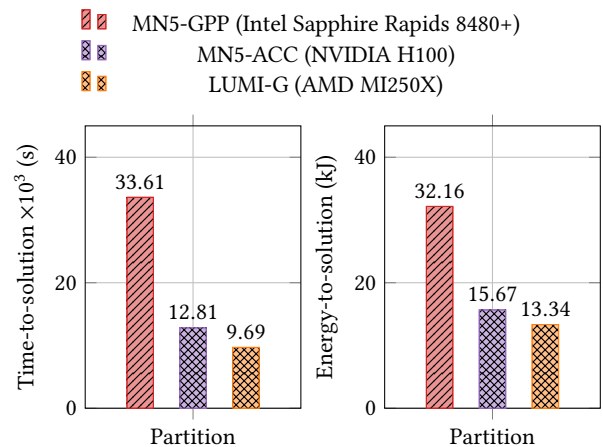
\begin{figure}
    \centering
    \begin{center}
        \ref{flexi_comp_leg2}
    \end{center}
    \begin{tikzpicture}
        \begin{axis}[
            ybar,
            ylabel={Time-to-solution $\times 10^3$ (s)},
            symbolic x coords={Partition},
            xtick=data,
            ymin=0,
            width=0.25\textwidth,
            height=0.3\textwidth,
            ymax=45,
            legend style={draw=none, legend columns=1, column sep=2ex},
            legend entries={MN5-GPP (Intel Sapphire Rapids 8480+), MN5-ACC (NVIDIA H100), LUMI-G (AMD MI250X)},
            legend to name=flexi_comp_leg2,
            nodes near coords,
            grid,
        ]
        \addplot[draw=matplotlibRed,fill=matplotlibRed!50, bar shift=-20pt, postaction={pattern=north east lines}] plot coordinates {
            (Partition, 33.61133)
        };
        \addplot[draw=matplotlibPurple,fill=matplotlibPurple!50, bar shift=0pt, postaction={pattern=crosshatch}] plot coordinates {
            (Partition, 12.80726)
        };
        \addplot[draw=matplotlibOrange,fill=matplotlibOrange!50, bar shift=20pt, postaction={pattern=crosshatch}] plot coordinates {
            (Partition, 9.69270)
        };
        \end{axis}
    \end{tikzpicture}
    \begin{tikzpicture}
        \begin{axis}[
            ybar,
            ylabel={Energy-to-solution (kJ)},
            symbolic x coords={Partition},
            xtick=data,
            ymin=0,
            ymax=45,
            nodes near coords,
            grid,
            width=0.25\textwidth,
            height=0.3\textwidth,
        ]
        \addplot[draw=matplotlibRed,fill=matplotlibRed!50, bar shift=-20pt, postaction={pattern=north east lines}] plot coordinates {
            (Partition, 32.16)
        };
        \addplot[draw=matplotlibPurple,fill=matplotlibPurple!50, bar shift=0pt, postaction={pattern=crosshatch}] plot coordinates {
            (Partition, 15.67)
        };
        \addplot[draw=matplotlibOrange,fill=matplotlibOrange!50, bar shift=20pt, postaction={pattern=crosshatch}] plot coordinates {
            (Partition, 13.34)
        };
        \end{axis}
    \end{tikzpicture}
    \caption{Time-to-solution (left) and energy-to-solution (right) on a single node for FLEXI/\galexi.}
\label{fig:flexi_energy}
\end{figure}

\section{Neko \& Nekbone}
\label{sec:neko}
Neko~\cite{neko} is a portable simulation framework based on high-order Spectral Element Method (SEM) on hexahedral meshes, mainly focusing on incompressible flow simulations. The framework is written in modern Fortran. It adopts an object-oriented approach, which allows for multi-tier abstractions of the solver stack and facilitates multiple hardware backends, ranging from general-purpose processors to accelerators and vector processors. 
Neko focuses on single core/ single accelerator efficiency via tensor product operator evaluations. A key to achieving good performance in spectral element methods is to consider a matrix-free formulation, where one always works with the unassembled matrix on a per-element basis.

Nekbone\cite{nekbone} is a mini-app that captures the fundamental design of Nek5000\cite{nek5000}, a large-scale, high-order solver for incompressible Navier-Stokes equations based on SEM. Nekbone solves a standard Poisson equation by partitioning the computational domain into high-order quadrilateral elements and using the Conjugate Gradient (CG) method as its main computational kernel. The CG solver can optionally be compiled with a multigrid preconditioner.

\subsection{Simulation Setup}
\textbf{Neko test case}: The simulation domain is a cubic region extending from $0$ to $8$ in each spatial direction, resulting in a computational domain of $8\times 8\times 8$. Within this domain, the Poisson equation $\nabla^2\varphi=f$ for pressure in the Navier-Stokes equation is solved using the Conjugate Gradient (PCG) method with the Jacobi preconditioner. The mesh comprises $16,384$ spectral elements with a polynomial order of $7$. The simulation is executed using 80 MPI ranks, corresponding to the total number of cores on a single node of MareNostrum5. This smaller-scale configuration is employed as a test case to explore and evaluate mixed-precision strategies within Neko.

\textbf{Nekbone test case}: Nekbone is also benchmarked by solving the Poisson equation. A multigrid preconditioner is used for the CG method. This configuration is set up as a weak scaling problem, where the problem size per MPI is constant (128 spectral elements with a polynomial order of $10$). The simulation is run using a total of 80 MPI ranks, resulting in a total problem size of $10,240$ elements (128 elements/rank $\times$ 80 ranks).

The insights gained from this mixed-precision exploration is being extended to the LHC6 simulation of a merchant ship hull, which requires the preconditioned Coupled Conjugate Gradient (PCG) method as the velocity solver and the Generalized Minimal Residual (GMRES) method as the pressure solver. The idea is to develop mixed-precision algorithmic solutions that interleave numerical robustness with efficiency, offering a novel perspective on algorithmic design aimed at optimizing both time-to-solution and energy-to-solution.





\subsection{Energy Consumption}
Building on the work of Chen et al. \cite{enablingmixedfgcs} reported an energy gain of 1.32x on the pressure solver with Neko, here we further explore the trade-off between energy savings and the tolerance for the solver. We conducted experiments of energy consumption by varying the PCG solver's convergence tolerance, sweeping from a loose ($tol=10^{-4}$) to a stringent ($tol=10^{-10}$) threshold.

Our experiments were conducted on MareNostrum5 (Intel Xeon Platinum 8460Y+) with energy data collected using the EAR software. To ensure accurate and low-noise measurements, we enforced exclusive node allocation using the \texttt{--exclusive} directive in our SLURM job scripts. The EAR plugin was enabled via \texttt{--ear=on} , and its policy is specified with \texttt{--ear-policy=monitoring}. Each test was executed $10$ times. Following job completion, we obtained the total energy consumption using the \texttt{eacct -j <JobID>} command (similar to \texttt{sacct}) and reported the median value.

\begin{figure}
    \centering
    \begin{tikzpicture}
    \begin{axis}[
        hide axis,
        xmin=0, xmax=1,
        ymin=0, ymax=1,
        legend style={
            draw=none,
            legend columns=2,
            column sep=2ex,
            /tikz/every even column/.append style={column sep=1em},
            at={(0.5,1.2)},
            anchor=south
        },
        legend to name=nekbone_legend
    ]
    \addlegendimage{mark=square, thick, color=matplotlibOrange}
    \addlegendentry{Mixed-precision}
    \addlegendimage{mark=triangle, thick, color=matplotlibBlue}
    \addlegendentry{Double-precision}
    \end{axis}
    \end{tikzpicture}

    \begin{center}
        \ref{nekbone_legend}
    \end{center}
    
    \begin{tikzpicture}
    \begin{axis}[
        width=8cm, height=4.5cm,
        xmode=log,
        log basis x=10,
        scaled y ticks = manual,
        yticklabel={\pgfmathparse{\tick/100}\pgfmathprintnumber{\pgfmathresult}},
        ytick scale label code/.code={$\times10^{2}$},
        xlabel={Tolerance},
        ylabel={Energy-to-solution (J)},
        xtick={1e-10,1e-9,1e-8,1e-7,1e-6,1e-5,1e-4},
        xticklabels={$10^{-10}$,$10^{-9}$,$10^{-8}$,$10^{-7}$,$10^{-6}$,$10^{-5}$,$10^{-4}$},
        grid=both,
    ]
    \addplot+[mark=square, thick, color=matplotlibOrange]
    coordinates {
    (1e-10,3857)
    (1e-9,3275)
    (1e-8,2906)
    (1e-7,2640)
    (1e-6,2551)
    (1e-5,2458)
    (1e-4,2297)
    };

    \addplot+[mark=triangle, thick, color=matplotlibBlue]
    coordinates {
    (1e-10,4920)
    (1e-9,3990)
    (1e-8,3204)
    (1e-7,3100)
    (1e-6,2760)
    (1e-5,2570)
    (1e-4,2402)
    };
    \end{axis}
    \end{tikzpicture}

    \vspace{0.1cm}
    
    \begin{tikzpicture}
    \definecolor{myblue}{RGB}{0,114,178}
    \begin{axis}[
        width=8cm, height=4.5cm,
        xmode=log,
        log basis x=10,
        xlabel={Tolerance},
        ylabel={Time-to-solution (s)},
        xtick={1e-10,1e-9,1e-8,1e-7,1e-6,1e-5,1e-4},
        xticklabels={$10^{-10}$,$10^{-9}$,$10^{-8}$,$10^{-7}$,$10^{-6}$,$10^{-5}$,$10^{-4}$},
        grid=both,
        legend style={draw=none, legend columns=2, column sep=2ex},
        legend entries={Mixed-precision, Double-precision},
        legend to name={nekbone_leg2}
    ]

    \addplot+[mark=square, thick, color=matplotlibOrange]
    coordinates {
    (1e-10,10.971)
    (1e-9,9.869)
    (1e-8,9.810)
    (1e-7,9.209)
    (1e-6,9.085)
    (1e-5,8.652)
    (1e-4,8.576)
    };

    \addplot+[mark=triangle, thick, color=matplotlibBlue]
    coordinates {
    (1e-10,12.090)
    (1e-9,10.612)
    (1e-8,10.306)
    (1e-7,10.104)
    (1e-6,9.725)
    (1e-5,9.326)
    (1e-4,9.232)
    };
    \end{axis}
    \end{tikzpicture}

    \caption{Energy-to-solution (top) and time-to-solution (bottom) vs. tolerance for mixed- and double-precision Nekbone (Poisson's equation) on a single node of MareNostrum5.}
    \label{fig:nekbone-tol}
\end{figure}
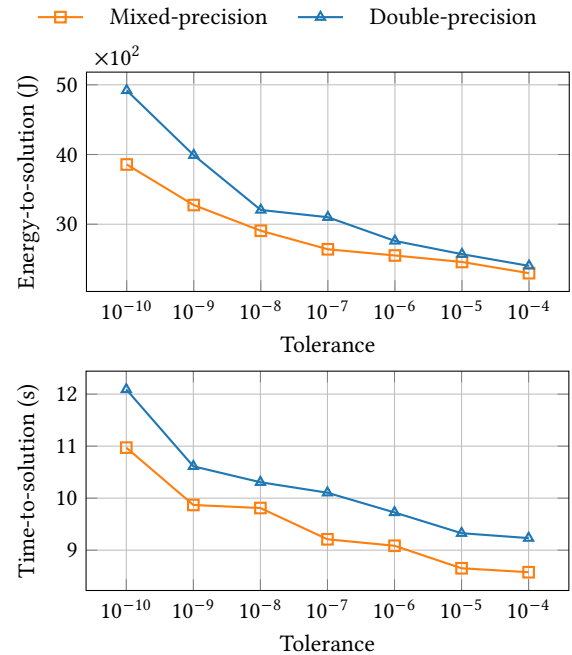

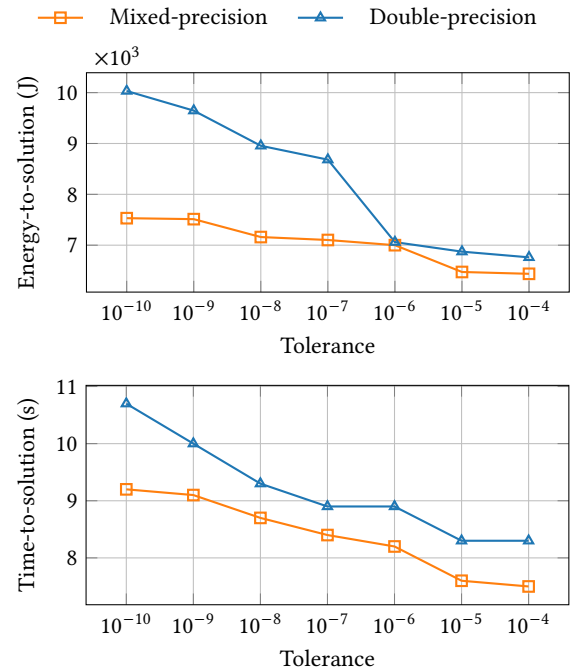
\begin{figure}
    \centering
    \begin{tikzpicture}
    \begin{axis}[
        hide axis,
        xmin=0, xmax=1,
        ymin=0, ymax=1,
        legend style={
            draw=none,
            legend columns=2,
            column sep=2ex,
            /tikz/every even column/.append style={column sep=1em},
            at={(0.5,1.2)},
            anchor=south
        },
        legend to name=neko_legend
    ]
    \addlegendimage{mark=square, thick, color=matplotlibOrange}
    \addlegendentry{Mixed-precision}
    \addlegendimage{mark=triangle, thick, color=matplotlibBlue}
    \addlegendentry{Double-precision}
    \end{axis}
    \end{tikzpicture}

    \begin{center}
        \ref{neko_legend}
    \end{center}
    
    \begin{tikzpicture}
    \begin{axis}[
        width=8cm, height=4.5cm,
        xmode=log,
        log basis x=10,
        scaled y ticks = manual,
        yticklabel={\pgfmathparse{\tick/1000}\pgfmathprintnumber{\pgfmathresult}},
        ytick scale label code/.code={$\times10^{3}$},
        xlabel={Tolerance},
        ylabel={Energy-to-solution (J)},
        xtick={1e-10,1e-9,1e-8,1e-7,1e-6,1e-5,1e-4},
        xticklabels={$10^{-10}$,$10^{-9}$,$10^{-8}$,$10^{-7}$,$10^{-6}$,$10^{-5}$,$10^{-4}$},
        grid=both
    ]
    \addplot+[mark=square, thick, color=matplotlibOrange]
    coordinates {
    (1e-10,7531)
    (1e-9,7512)
    (1e-8,7160)
    (1e-7,7102)
    (1e-6,7000)
    (1e-5,6473)
    (1e-4,6437)
    };

    \addplot+[mark=triangle, thick, color=matplotlibBlue]
    coordinates {
    (1e-10,10032)
    (1e-9,9646)
    (1e-8,8955)
    (1e-7,8682)
    (1e-6,7060)
    (1e-5,6873)
    (1e-4,6761)
    };
    \end{axis}
    \end{tikzpicture}

    \vspace{0.1cm}
    \begin{tikzpicture}
    \definecolor{myblue}{RGB}{0,114,178}
    \begin{axis}[
        width=8cm, height=4.5cm,
        xmode=log,
        log basis x=10,
        xlabel={Tolerance},
        ylabel={Time-to-solution (s)},
        xtick={1e-10,1e-9,1e-8,1e-7,1e-6,1e-5,1e-4},
        xticklabels={$10^{-10}$,$10^{-9}$,$10^{-8}$,$10^{-7}$,$10^{-6}$,$10^{-5}$,$10^{-4}$},
        grid=both
    ]

    \addplot+[mark=square, thick, color=matplotlibOrange]
    coordinates {
    (1e-10,9.2)
    (1e-9,9.1)
    (1e-8,8.7)
    (1e-7,8.4)
    (1e-6,8.2)
    (1e-5,7.6)
    (1e-4,7.5)
    };

    \addplot+[mark=triangle, thick, color=matplotlibBlue]
    coordinates {
    (1e-10,10.7)
    (1e-9,10)
    (1e-8,9.3)
    (1e-7,8.9)
    (1e-6,8.9)
    (1e-5,8.3)
    (1e-4,8.3)
    };
    \end{axis}
    \end{tikzpicture}

    \caption{Energy-to-solution (top) and time-to-solution (bottom) vs. tolerance for mixed-precision and double-precision Neko (Poisson's equation) on a single node of MareNostrum5.}
    \label{fig:neko-tol}
\end{figure}

\cref{fig:nekbone-tol} and \cref{fig:neko-tol} illustrate the energy-to-solution (top) and time-to-solution (bottom), comparing the mixed-precision and double-precision implementations across a range of PCG tolerances. We observe two key points:

\begin{itemize}
    \item \textit{Mixed-precision advantage}: The mixed-precision implementation clearly outperforms the double-precision baseline. It consistently achieves a lower time-to-solution and consumes less energy.
    \item \textit{Impact of tolerance}: Both mixed-precision and double-precision demonstrate that relaxing the solver tolerance leads to a general decrease in both time and energy-to-solution, as fewer iterations are required to reach convergence. This benefit has a potential to flourish when combined with the discretization error and mesh refinement.
\end{itemize}

\section{NekRS}
\label{sec:nekrs}

During the CEEC project, the GPU-accelerated high-order SEM code NekRS~\cite{fischer2021nekrs} was  employed to perform ultra-high-resolution Large Eddy Simulations (LES) for investigating both stable and convective (unstable) Atmospheric Boundary Layer (ABL) dynamics, collectively referred to as LHC5. For validation and model inter-comparison, the Global Energy and Water Cycle Experiment (GEWEX) Atmospheric Boundary Layer Study (GABLS)~\cite{GABLS} benchmark was used to represent the stably stratified ABL. The study was to assess the sensitivity of LES solutions to grid resolution, subgrid-scale (SGS) model parameters, numerical discretization schemes, and surface boundary conditions. 

To fully exploit heterogeneous architectures, specialized OCCA-based computational kernels were developed for NekRS, optimizing memory hierarchy and minimizing data transfer between host and device. Scalability to millions of MPI ranks or GPUs is achieved through the gsLib communication library, which enables efficient MPI-based gather–scatter operations, rapid near-neighbor/halo exchanges, and highly optimized stencil communications for the lower levels of the multigrid pressure solver. Domain partitioning is performed using parallel recursive spectral bisection (parRSB) to ensure a balanced distribution of elements across processors. More information on numerical implementation can be found in \cite{fischer2021nekrs}\cite{Fischer2005Hybrid}\cite{Phillips2021Tuning}.

\subsection{Simulation Setup}
For the GABLS setup, the Reynolds number (Re) based on the reference length, velocity scale and molecular viscosity, exceeds 50 millions, which precludes direct numerical simulation (DNS) wherein all turbulent scales are resolved. Consequently, LES were performed in a box domain ($400\times400\times400\,\,m^3$) by considering various SGS models and various resolutions up to $2048^3$ total grid points. At the bottom boundary representing the lower wall, the BCs are based on the Monin-Obukhov similarity theory, where a zero normal velocity is imposed together with traction BCs for the two horizontal velocity components as well as heat flux BCs for the potential temperature. The traction BCs are based on the horizontally-averaged slip velocity along the lower boundary. 

\subsection{Energy Consumption}
Regarding energy consumption for NekRS, the Jülich LLview~\cite{LLview2023}
profiling tool was utilized, see~\cref{sec:subsec:subsubsec-LLview}. To estimate the energy
consumption of the simulations, the average power consumed by
each GPU measured by LLView was multiplied by the number of
GPUs utilized and the total walltime of the simulation.

\begin{figure}
    \centering

    \begin{tikzpicture}
    \definecolor{blue}{RGB}{76,114,176}
    \definecolor{green}{RGB}{85,168,104}
    \definecolor{red}{RGB}{196,78,82}

    \begin{axis}[
        ybar,
        bar width=8pt,
        bar shift=-8pt, 
        axis line style={gray},
        tick pos=left,
        x grid style={gray!50},
        y grid style={gray!50},
        xlabel={{\# Devices}},
        ylabel={{Energy Consumption [kWh]}},
        ymin=0, ymax=100,
        xtick={0,1,2,3,4,5,6,7,8,9},
       xticklabels={24,48,96,192,288,384,760,1008,1540,2040},
       ytick={0,20,40,60,80,100},
       width=9cm, height=6cm,
       ymajorgrids,
       legend style={at={(0.02,0.98)},anchor=north west, draw=gray},
       nodes near coords,
       every node near coord/.append style={font=\small, yshift=2pt},
       legend image code/.code={%
            \draw[#1,fill=#1,draw=gray] (0cm,-0.1cm) rectangle (0.3cm,0.1cm);
        },
    ]

    \addplot[ybar,draw=matplotlibRed,fill=matplotlibRed!50, bar shift=-8pt, postaction={pattern=north east lines}] coordinates {
    (0.4,0.8) (1.4,0.92) (2.2,0.97)
    };
    \addlegendentry{$512^3$}

    \addplot[ybar,draw=matplotlibPurple,fill=matplotlibPurple!50, bar shift=0pt, postaction={pattern=crosshatch}] coordinates {
    (2.2,8.1) (3,7.93) (4,8.32) (5,9.32)
    };
    \addlegendentry{$1024^3$}

    \addplot[ybar,draw=matplotlibOrange,fill=matplotlibOrange!50, bar shift=8pt, postaction={pattern=crosshatch}] coordinates {
    (5.6,70) (6.6,72) (7.6,76) (8.6,82)
    };
    \addlegendentry{$2048^3$}

    \end{axis}
    \end{tikzpicture}

    \caption{Energy consumption per 5,000 timesteps for the GABLS case, NekRS with resolutions of $512^{3}$ , $1024^{3}$ and $2048^{3}$ on Nvidia A100 GPUs (JUWELS Booster).}
    \label{fig:NekRS_energy_consumption}
\end{figure}
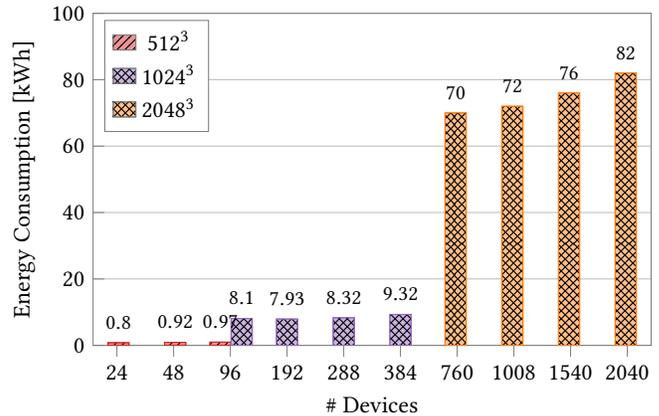

\cref{fig:NekRS_energy_consumption}  presents the energy consumption in kilowatt-hours (kWh) per 5000 timesteps for simulations with resolutions of $512^3$, $1024^3$, and $2048^3$ gridpoints using NVIDIA A100 GPUs across varying numbers of GPUs. Each increment in resolution increases DoFs by a factor of eight compared to the previous resolution. The figure illustrates that the energy consumption scales approximately eight-fold with each increase in resolution, which is the same as the factor of increase of DoFs.

\begin{figure}
    \centering
    \begin{tikzpicture}
    
    \definecolor{matplotlibRed}{RGB}{196,78,82}
    \definecolor{matplotlibPurple}{RGB}{129,114,179}
    \definecolor{matplotlibOrange}{RGB}{255,128,14}

    \begin{axis}[
        width=9cm, height=6cm,
        xlabel={{\# Devices}},
        ylabel={{PE \hspace{0.05cm} \& \hspace{0.05cm} Energy ratio [-]}},
        xmode=log,
        xmin=20, xmax=2200,
        ymin=0.65, ymax=1.35,
        ytick={0.7,0.8,0.9,1.0,1.1,1.2,1.3},
        tick pos=left,
        axis line style={gray},
        grid=both,
        grid style={gray!50,solid},
        minor grid style={gray!50,dotted},
        legend style={
            at={(0.01,0.96)},
            anchor=north west,
            font=\scriptsize,
            draw=gray,
            fill=white,
            fill opacity=0.9,
            legend columns=1,
        },
    ]

    \addplot[draw=matplotlibRed,mark=o,very thick] coordinates {(24,1.0) (48,0.88) (96,0.7)};
    \addplot[draw=matplotlibPurple,mark=o,very thick] coordinates {(96,1.0) (192,0.945) (288,0.815) (384,0.755)};
    \addplot[draw=matplotlibOrange,mark=o,very thick] coordinates {(760,1.0) (1008,0.94) (1540,0.875) (2040,0.72)};

    \addplot[draw=matplotlibRed,mark=s,very thick,dashed] coordinates {(24,1.0) (48,1.06) (96,1.08)};
    \addplot[draw=matplotlibPurple,mark=s,very thick,dashed] coordinates {(96,1.0) (192,0.986) (288,1.04) (384,1.13)};
    \addplot[draw=matplotlibOrange,mark=s,very thick,dashed] coordinates {(760,1.0) (1008,1.002) (1540,1.05) (2040,1.13)};

    \addlegendimage{draw=matplotlibRed,very thick}  
    \addlegendentry{$512^3$}
    \addlegendimage{draw=matplotlibPurple,very thick} 
    \addlegendentry{$1024^3$}
    \addlegendimage{draw=matplotlibOrange,very thick} 
    \addlegendentry{$2048^3$}

    \end{axis}

    \node[anchor=south, yshift=-0.5cm,xshift=0.8cm, font=\scriptsize, draw=gray, fill=white, fill opacity=0.9, inner sep=3pt] at (current axis.north) {%
    \begin{tikzpicture}
        \draw[black,very thick] (0,0) -- (0.4,0) node[right, xshift=0.3em]{Parallel Efficiency};
        \draw[black,very thick,dashed] (2.4,0) -- (2.8,0) node[right, xshift=0.3em]{Energy Ratio};
    \end{tikzpicture}%
    };

    \end{tikzpicture}

    \caption{Parallel efficiency (PE) and energy ratio for the GABLS case, NekRS with resolutions of $512^{3}$ , $1024^{3}$ and $2048^{3}$ on Nvidia A100 GPUs (JUWELS Booster).}
    \label{fig:NekRS_energy_efficiency}
\end{figure}
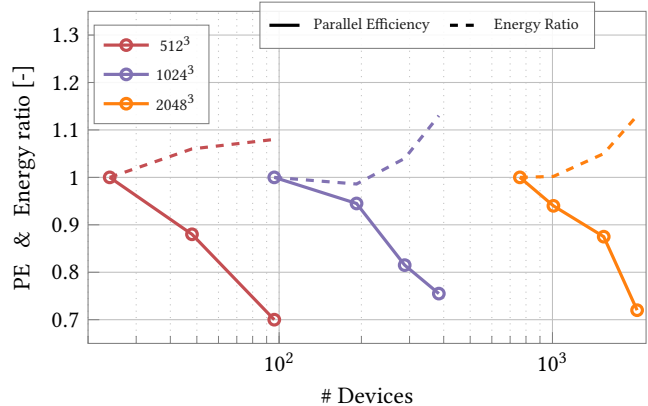

~\cref{fig:NekRS_energy_efficiency} depicts the parallel efficiency in relation to the energy ratio. As parallel efficiency decreases, the energy required to perform the same simulation also increases, since GPUs are underutilized. This, along with the fact that NekRS can utilize 80-90\% of the realizable peak memory bandwidth for the most communication-intensive kernels across multiple platforms while sustaining 8-10 TFLOPs (80-90\% of the peak computational power of the most advanced GPUs like Nvidia A100) for the most computation-intensive kernels, suggests that NekRS’s parallel and energy efficiency is bandwidth-limited and constrained by communication overhead rather than by hardware (computational power) when a large number of GPUs is utilized, specifically when n/P drops below 2.5 million.

\section{Concluding Remarks}
\label{sec:conclusions}
The work presented through the CEEC project highlights the imperative of fostering community-wide awareness and engagement in energy measurements within HPC and application communities. Despite the availability of advanced tools and methodologies, the lack of standardized and easily accessible mechanisms for energy monitoring continues to hinder consistent evaluation and comparison across different hardware and software architectures. 
Facilitating such measurements must therefore become a key objective to enable users and developers to make informed decisions that advance energy-efficient practices.

Porting of CFD applications to GPUs, one of the goals of the CEEC project, demonstrated to exhibit superiority in achieving favorable energy-to-solution metrics compared to CPU-based systems. 
The application case studies within CEEC highlighted that our codes are well optimized and demonstrate superior scalability. However, the results also show that underutilization of computational resources can have detrimental effects on both performance and energy consumption. 
Furthermore, the adoption of mixed-precision techniques has been shown to provide an effective balance between computational accuracy and energy efficiency, representing a promising direction for sustainable exascale applications.
The findings emphasize that optimization should not be limited to runtime reduction alone, but must equally consider the compute/storage precision and energy implications of numerical and architectural choices.


Overall, the findings highlight that sustainable advancements in HPC have to be achieved through an integrated approach that combines energy awareness, precision control, and resource-efficient computation. Establishing community standards and promoting transparent access to energy measurement data are essential next steps toward realizing energy-conscious exascale computing across Europe and beyond.


\begin{acks}
This work was funded by the European Union.
This work has received funding from the European High Performance Computing Joint Undertaking (JU) and Sweden, Germany, Spain, Greece, and Denmark under grant agreement No 101093393.
We acknowledge EuroHPC Joint Undertaking for awarding us access to LUMI at CSC, Finland, to MareNostrum5 as BSC, Spain, and to MeluXina at LuxProvide, Luxembourg.
\end{acks}

\printbibliography


\clearpage
\end{document}